# Threefold Analysis of Distributed Systems: IMDS, Petri Net and Distributed Automata DA$^3$

Wiktor B. Daszczuk
Institute of Computer Science,
Warsaw University of Technology
Nowowiejska Str. 15/19
00-665 Warsaw, Poland
Email: wbd@ii.pw.edu.pl

*Abstract*—Integrated Model of Distributed Systems is used for specification and verification of distributed systems. In the formalism, a system is modeled as a set of servers' states and agents' messages. The operation of a system is modeled as actions converting global system configuration (a set of states and messages) to a new configuration. The formalism is used in Dedan verification environment, in which specification and verification of distributed systems is performed. Equivalent Petri nets are used for structural analysis. For the graphical specification and simulation of distributed systems, Distributed Autonomous and Asynchronous Automata (DA$^3$) are invented. Such simulation does not require calculation of global configuration space of a system. Two forms of DA$^3$ are shown: Server-DA$^3$ (SDA$^3$) for the server view and Agent-DA$^3$ (ADA$^3$) for the agent view.

## I. INTRODUCTION

INTEGRATED Model of Distributed Systems (IMDS [1] [2]) is a formalism for specification of distributed systems. The formalism exploits the natural features of such systems:

- *Communication duality*: a model of a distributed system is uniform, it can be decomposed ("cut") to server processes communicating by messages or to agent processes communicating by shared resources (servers' states).
- *Locality of actions*: an action is executed entirely inside a server, on a basis of the server's state and a set of messages pending at this server.
- *Autonomy of decisions*: a server itself decides, which action will be executed (among enabled ones) and when.
- Asynchrony of actions: a server in given state awaits messages that enable some actions in it, or messages pend at a server, waiting for state that enables the server's actions.
- *Asynchrony of communication*: unidirectional cannels for message passing are assumed, without any acknowledgments (an acknowledgment may be sent, but using another asynchronous channel in opposite direction)

It is opposed to synchronous models, like CSP [3] or CCS [4], where distributed elements must agree on communication, which requires some kind of nonlocality.

The formal definition of IMDS can be found in [2], here we give an overview. IMDS is based on four sets: servers $S$, agents $A$, values $V$ and services $R$. We will use lowercase letters for elements of these sets. Any server has a set of *states* being pairs $p=(s,v)$. A server runs by accepting *messages* that invoke its *services*. The acceptance of a message at given state of a server causes the execution of an *action* in the context of this server. The action changes the state of the server and causes a next message to be sent (typically to some other server, but sometimes to the same server). In the system, distributed computations may be performed as sequences of actions. An *agent* is introduced to identify such a distributed computation.

The communication between servers is performed by means of messages. A message is an invocation of a server's service in a context of an agent, thus it is a triple: $m=(a,s,r)$. An action is defined for a pair $(m,p)$ and produces a new pair $(m',p')$. Therefore, the actions form a relation $(m,p)\lambda(m',p')$. We say that the pair $(m,p)$ matches if an action is defined for it. We impose the constraints on an action:

- The action is invoked by a message issued to a server and it is executed on this server, therefore server component ($s$) of the message $m$ and the state $p$ must be the same.
- The action produces a new state of the server, therefore the server component ($s$) of the input state $p$ and the output state $p'$, must be the same.
- The action produces a new message of the agent (a distributed computation), therefore the agent component ($a$) of the input message $m$ and the output message $m'$ must be the same.

A special kind of action terminates an agent, it is a relation between a pair $(m,p)$ and a singleton $(p')$ – the output message is missing: $(m,p)\lambda(p')$.

A global system configuration $T$ consists of current states of all servers and pending messages of agents (all but terminated ones). The actions are executed in interleaving way, therefore an action transforms its input configuration $T$, containing its input items $m$ and $p$, into an output configuration $T'$, containing its output items $m'$ and $p'$ (or $p'$ only). The initial configuration $T_0$ consists of initial states of all servers and initial messages of all agents.

The semantics of a distributed system is defined as Labeled Transition System (LTS), in which nodes are the configurations, initial node is initial configuration and transitions are actions.

A system may be decomposed into processes. If we group all actions to subsets on individual servers – we get *server processes*. Their carriers are servers' states and they communicate by messages. It is the *server view* of the system.

If we group all actions to subsets appointed to individual agents – we get *agent processes*. Their carriers are agents' messages and they communicate by servers' states (treated as resources). It is the *agent view* of the system.

A temporal logic is defined over IMDS, which allows to verify the correctness of modeled systems by model checking [5]. Especially, partial deadlock and termination (where not all processes participate) may be identified using universal temporal formulas. Universality of the formulas do not require from the designer any knowledge about temporal logic and model checking. Automatic partial deadlock and partial termination checking is a unique feature for systems without restriction on their structure (cycling, terminating, etc.).

The deadlocks identified using IMDS in the server view concern communication, while those detected in the agent view are resource deadlocks.

The formalism is incorporated in Dedan program for specification and verification of distributed systems [6]. Although automatic deadlock detection is useful, it has a typical drawback: model checking interrupts the verification if the evaluated formula is decided to be true or false. In deadlock detection, first deadlock found finishes the evaluation. Next deadlock may be identified after modification of the system. If many deadlocks are present, the procedure complicates the verification.

The situation is cured by static analysis of a Petri net corresponding to IMDS model [7]. In the network, subnets called *siphons* may be identified [8]: a siphon which is emptied from tokens, cannot restore them. If the emptying of a siphon is reachable, it may denote a deadlock. Also, some other properties of verified system may be identified using Petri net approach: unreachable actions ("dead code"), unrelated components, etc. A conversion of IMDS to a Petri net is described later in this paper.

In addition to static analysis (model checking and Petri net analysis), graphical tools for specification and simulation of the distributes systems are needed. The automata are widely used for such purpose, often called *distributed automata*. For example, automata on distributed alphabets, communicating on common symbols, based on Zielonka's automata [9]. These automata are called *distributed automata* in many papers concerning the behavior of concurrent systems (in some of them additionally equipped with real time clocks for temporal analysis with real-time constraints): [10] [11] [12]. These automata are called asynchronous for example in [13], although they perform actions (make the transitions) asynchronously only if the input symbols are distinct. They make synchronous moves on common input symbols (and it is the only common aspect of the automata). Similar are Timed Automata [10] and CSP processes [3], synchronizing on ! and ? operations rather than on symbols of input alphabet. Close to Zielonka's automata are Büchi automata. They differ in distinguishing some states as accepting. They are called distributed automata in [14].

Message Passing Automata (MPA, called distributed automata in [15]) are really distributed and asynchronous. They have ordered sets on symbols waiting for acceptance, called *buffers* or *queues*. Pushdown Distributed Automata (PDA) are equipped with local memories of input symbols (stacks) [16].

We developed a formalism of Distributed Autonomous and Asynchronous Automata (DAAA, $DA^3$ in short) for the graphic specification and simulation.

The three modeling methods: IMDS, corresponding Petri nets and Distributed Automata are equivalent. The generate reachability spaces which have identical structures. Therefore, they all highlight the natural features of distributed systems: communication duality, locality of actions, autonomy of decisions, asynchrony of actions and communication. This paper describes the application of the three formalisms to the verification of distributed systems.

A Dedan program is presented in Section II. The example of a bounded buffer specification in IMDS is given in Section III. The conversion of IMDS to Petri net is described in Section IV. The notion of $DA^3$ distributed automata is covered in detail in Section V. Two versions – Server $DA^3$ ($SDA^3$) and Agent $DA^3$ ($ADA^3$) are defined. The operation of Dedan program on $DA^3$ is described in Section VI. Conclusions and further work are covered in Section VII.

## II. THE DEDAN PROGRAM

The IMDS formalism was used, together with model checking technique [5], to develop the Dedan program which finds various kinds of deadlock and termination in a verified system. A counterexample is generated if a deadlock is found, or it is a witness of distributed termination. Also, observation of global reachability graph and simulation over this graph are possible.

Dedan is built in such a way that the specification of temporal formulas and temporal verification are hidden to a user. The reason is that model checking techniques are seldom known by the engineers. Therefore, the program is constructed in such a way that a user specifies the system and simply "pushes the button" to check for the existence of deadlocks or to check distributed termination.

Although the main target of Dedan is finding deadlocks and termination checking, a user may be interested in other properties of a verified system, for example:

- structural properties of a system: structural conflicts, dead code, pure cyclic system or not, etc.,
- temporal properties other than deadlock and termination: if a system is safe from some erroneous situation, if given situations are inevitable, etc.,
- graphical definition of concurrent components of a system (servers or agents),
- graphical simulation in terms of concurrent components rather than in terms of a global graph.

For the purpose of supporting the above possibilities, some additional facilities are included into Dedan:

- export to external model checkers for temporal analysis, for example Uppaal [17],
- export of a model to Petri net form for the analysis under Charlie Petri net analyzer [18] – to obtain a structural analysis,
- alternative formulation of a system as Distributed Automata, for facilitation of a system specification and simulation in terms of distributed components – automata representing server processes or agent processes.

### III. SIMPLE EXAMPLE – BUFFER

To present the two views of IMDS model, a simple system containing a buffer with producer and consumer agents (each one originating from its own server) is included below. First the server view follows. The notation is intuitional: server types are defined (lines 2, 9, 16, (formal parameters specify agents and other servers used)). Every server includes states (l.3, 10, 17), services (l.4, 11, 18) and actions (l.6-7, 13-14, 20-21) (an action $((a,s,r),(s,v))\lambda((a,s',r'),(s,v'))$ has the form {a.s.r, s.v}→{a.s'.r', s.v'}). Then, server and agent variables are declared (l.23-24). The variables have the same names as the types, they are distinguished by context. If a variable has the same identifier as its type, a declaration *variable:type* may be suppressed to a single identifier, as in the example. At the end, servers (l.26-28) and agents (l.29-30) are initialized (and variable names are bound with formal parameters of servers).

1. **system** BUF_server_view;

2. **server**: buf (**agents** Aprod,Acons; **servers** Sprod,Scons),
3. **services** {put, get},
4. **states** {no_elem,elem},
5. **actions** {
6.   {Aprod.buf.put, buf.no_elem} ->
        {Aprod.Sprod.ok_put, buf.elem},
7.   {Acons.buf.get, buf.elem} ->
        {Acons.Scons.ok_get, buf.no_elem},
8. }

9. **server**: Sprod (**agents** Aprod; **servers** buf),
10. **services** {doSth,ok_put}
11. **states** {neutral,prod}
12. **actions** {
13.   {Aprod.Sprod.doSth, Sprod.neutral} ->
         {Aprod.buf.put, Sprod.prod}
14.   {Aprod.Sprod.ok_put, Sprod.prod} ->
         {Aprod.Sprod.doSth, Sprod.neutral}
15. }

16. **server**: Scons (**agents** Acons; **servers** buf),
17. **services** {doSth,ok_get}
18. **states** {neutral,cons}
19. **actions** {
20.   {Acons.Scons.doSth, Scons.neutral} ->
         {Acons.buf.get, Scons.cons}
21.   {Acons.Scons.ok_get, Scons.cons} ->
         {Acons.Scons.doSth, Scons.neutral}
22. }

23. **servers** buf,Sprod,Scons;
24. **agents** Aprod,Acons;

25. **init** -> {
26.   Sprod(Aprod,buf).neutral,
27.   Scons(Acons,buf).neutral,
28.   buf(Aprod,Acons,Sprod,Scons).no_elem,

29.   Aprod.Sprod.doSth,
30.   Acons.Scons.doSth,
}.

The Dedan program automatically converts the specification to the agent view. Now, the actions are grouped in agents rather than in servers (lines 13-15, 19-21).

1. **system** BUF_agent_view;

2. **server**: buf,
3. **services** {put, get}
4. **states** {no_elem, elem};

5. **server**: Sprod,
6. **services** {doSth, ok_put}
7. **states** {neutral, prod};

8. **server**: Scons,
9. **services** {doSth, ok_get}
10. **states** {neutral, cons};

11. **agent**: Aprod (**servers** buf:buf,Sprod:Sprod),
12. **actions** {
13.   {Aprod.buf.put, buf.no_elem} ->
         {Aprod.Sprod.ok_put, buf.elem},

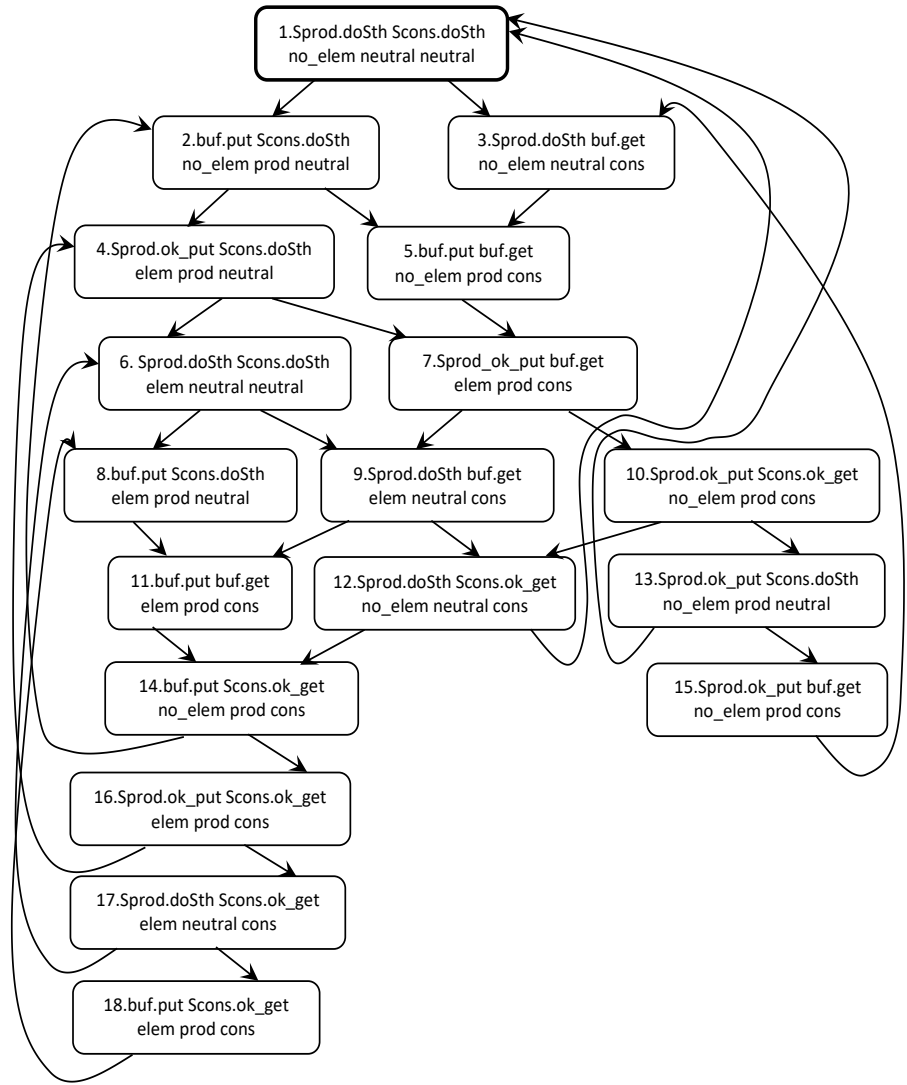

Fig. 1. LTS for the *buffer* system

14. {Aprod.Sprod.doSth, Sprod.neutral} ->
    {Aprod.buf.put, Sprod.prod},
15. {Aprod.Sprod.ok_put, Sprod.prod} ->
    {Aprod.Sprod.doSth, Sprod.neutral},
16. };

17. **agent**: Acons (**servers** buf:buf,Scons:Scons),
18. **actions** {
19. {Acons.buf.get, buf.elem} ->
    {Acons.Scons.ok_get, buf.no_elem},
20. {Acons.Scons.ok_get, Scons.cons} ->
    {Acons.Scons.doSth, Scons.neutral},
21. {Acons.Scons.doSth, Scons.neutral} ->
    {Acons.buf.get, Scons.cons},
22. };

23. **agents**   Aprod:Aprod,Acons:Acons;
24. **servers**  buf:buf,Sprod:Sprod,Scons:Scons;

25. **init** -> {
26. Aprod(buf,Sprod).Sprod.doSth,
27. Acons(buf,Scons).Scons.doSth,

28. buf.no_elem,
29. Sprod.neutral,
30. Scons.neutral,
31. }

The LTS of the example system is presented in Fig. 1. In the nodes, messages of the agents *Sprod* and *Scons* are displayed (without agent identifiers) in the first line and the states of all servers (*buf, Sprod, Scons*) are displayed in the second line (without server identifiers). Of course, this LTS generated both from the server view and from the agent view is identical, as the views are projections onto servers and onto agents of a uniform system.

## IV. PETRI NET EQUIVALENT TO IMDS

A designer may be interested in some structural properties of a verified system, for example structural conflicts, dead code, pure cyclic system or not, etc. For this purpose, Petri nets equivalent to IMDS models were elaborated. The Dedan program exports IMDS models to Charlie Petri net analyzer [19] [18]. The export is in ANDL format (*Abstract Net Description Language* [20]).

An IMDS system is converted to an equivalent Petri net in such a way that every action is converted to a Petri net transition, as illustrated in Fig. 2. Input items (a message *m* and a state *p*) are converted to the input places *m* and *p*. In a regular action, output items (a message *m'* and a state *p'*) are converted to the output places *m'* and *p'* (Fig. 2a). In an agent-terminating action, only one output place is present (corresponding to an output state *p'*, Fig. 2b). The initial marking of the Petri net has tokens in all places of initial servers' states and all places of initial agents' messages. By construction of the described conversion of an IMDS system to a Petri net, the reachable markings graph has identical structure as LTS of IMDS (states ↔ "state" places, messages ↔ "message" places, actions ↔ transitions, configuration ↔ marking, initial configuration ↔ initial marking).

The Petri net is not colored in the sense of [21], but we use read filling for states and green filling for messages in all figures, for readability.

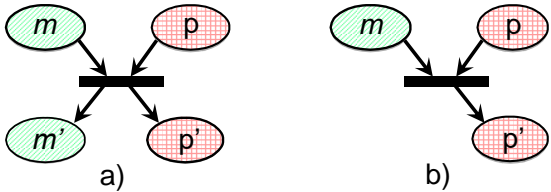

Fig. 2 Petri net interpretation of a) regular action
b) agent-terminating action

The Petri net of a *buffer* system is illustrated in Fig. 3. The states and messages in individual servers are grouped and separated by dashed lines. The states of servers are filled red while the messages are filled green. Also, `Sprod` states have dense grill while `Scons` states have rare grill. States of `buf` have chessboard filling. Messages of `Aprod` have diagonal hatching while messages of `Acons` have horizontal hatching. Initial states and initial messages are surrounded by bold ovals. All messages have identifiers in italics.

## V. DISTRIBUTED AUTONOMOUS AND ASYNCHRONOUS AUTOMATA ($DA^3$)

In computer engineering practice, various forms of automata are used to express the behavior of concurrent components. There are two reasons: graphical representation and individual modeling of distinct components. UML state diagrams are the good example [22].

For a graphical representation of distributed systems, and for a simulation in terms of parallel components of a system, Distributed Autonomous and Asynchronous Automata ($DA^3$) were invented. We claim that our distributed automata are better to describe parallelism and cooperation in real distributed environment (with full asynchrony) than other formalisms in the literature known as *distributed automata*, mentioned is Section I.

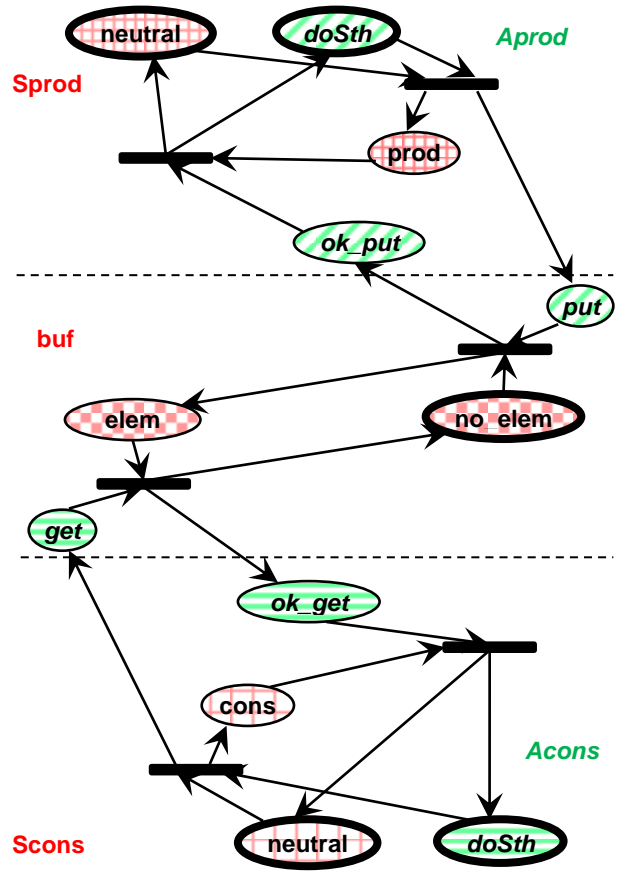

Fig. 3 Petri net representation of the *buffer* system:
servers `Sprod, Scons, buf`, agents `Aprod, Acons`

We introduce a new version of distributed automata, equivalent to IMDS formalism. We call them Distributed Autonomous, Asynchronous Automata – $DA^3$ (D-tripleA or DA-cubed) to distinguish them from all the previously mentioned formalisms, all called *distributed automata*. Our automata reflect the behavior of distributed components. The servers make decisions (perform actions) individually without any knowledge of other servers (autonomy) and messages are sent regardless of the states of target servers (asynchrony). As there are two views of a distributed system in IMDS, two forms of $DA^3$ were developed – Server-$DA^3$ and Agent-$DA^3$ ($SDA^3$ and $ADA^3$).

### A. Server automata ($SDA^3$)

An IMDS system in the server view may be shown as a set of communicating automata $SDA^3$ (Distributed Server Automata), similar to MPA:

- States of a server are nodes (we use *node* instead of *state* to avoid ambiguity) of corresponding automaton.

- An initial state of the server is an initial node of the automaton.
- Actions of the server process are transitions of the automaton.
- The automaton is Mealy-style [23], labels of the transitions in the automaton have the form extracted from actions; an IMDS action $(m,p)\lambda(m',p')$ is converted to a transition from $p$ to $p'$ with a label $m/m'$. It is a triple (*node, transition label, node*): $(p,m/m',p')$.
- The automaton is equipped with an input set – a set of input symbols pending, corresponding to a set of pending messages at the server. Firing a transition $(p,m/m',p')$ in the automaton of server $s$ retrieves the symbol $m$ from the input set of this automaton and inserts the symbol $m'$ to the input set of an automaton of the server $s'$ appointed by $m'$. An initial input set consists of initial messages of agents directed to this server.
- The special agent-terminating action $(m,p)\lambda(p')$ is converted to a transition that does not produce an output symbol: $(p,m/,p')$.

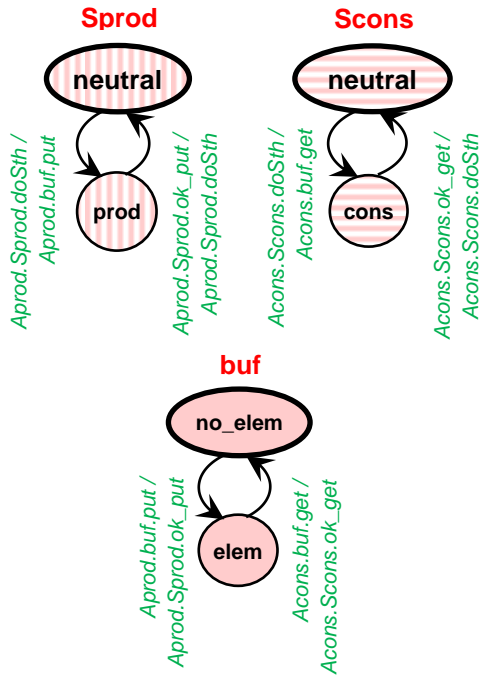

Fig. 4 Server automata of the *buffer* system

We denote automaton corresponding to a server $s$ as $\mathscr{z}$ (reflected $s$). The set of all server automata we denote as $\mathscr{Z}$. To define a behavior of server automata, we define a *position* of an automaton $(p,X)$ which consists of a node $p$ and given value (content) of the input set $X$. A *global position* of $\mathscr{Z}$ is a set of nodes of all server automata (servers' states) and the union of their current input sets.

The execution of a transition in a server automaton moves its current node to the target node of the transition, extracts an input message from the input set of the automaton and inserts the output message to the input set of the automaton appointed by the output message. As in IMDS, the transitions are executed in interleaving manner, and in a case of many transitions possible, the choice is nondeterministic.

The server automata of the *buffer* system are presented in Fig. 4. A *global graph* of $\mathscr{Z}$ cooperation may be elaborated in such a way that nodes are global positions, and edges are transitions in individual server automata. Of course, this graph is analogous to the LTS of IMDS system.

The initial sates of servers in Fig. 4 are in bold ovals. Server names are omitted in the state labels, because they are identical for all states in given server automaton.

Every automaton is equipped with the input set of pending messages, not shown in Fig. 4:
$X_{buf} \subseteq exp(\{$ `(Aprod,buf,put)`,
  `(Acons,buf,get)` $\})$,
$X_{Sprod} \subseteq exp(\{$ `(Aprod,Sprod,doSth)`,
  `(Aprod,Sprod,ok_put)` $\})$,
$X_{Scons} \subseteq exp(\{$ `(Acons,Scons,doSth)`,
  `(Acons,Scons,ok_get)` $\})$.
The initial input sets are:
$X_{0\ buf} = \emptyset$,
$X_{0\ Sprod} = \{$ `(Aprod,Sprod,doSth)` $\}$,
$X_{0\ Scons} = \{$ `(Acons,Scons,doSth)` $\}$.
The SDA$^3$ are similar to Message Passing Automata. The difference is in the ordering of messages on the input of the automaton: in MPA pending messages are ordered in the input queue (or input buffer) [15], while in SDA$^3$ any message form the input set may cause a transition (no ordering). If the input buffers of MPA are bounded, a deadlock may occur because of all processes sending to full buffers. Such a situation occurs when the size of buffers is taken too small. IMDS helps to overcome this problem by posing an accurate limit for the input set maximum size (or the input buffer): it is simply the number of agents.

*B. Agent automata (ADA$^3$)*

An IMDS system in the agent view may be shown as a set of communicating automata ADA$^3$ (Agent Distributed Autonomous and Asynchronous Automata). We use term *node* in these automata instead of *state*, because states ate attributed to servers in IMDS and it may be misleading. The ADA$^3$ automata are similar to Timed Automata with variables used in Uppaal [17] (but we consider only timeless systems here):

- Messages of an agent are *nodes* of a corresponding automaton.
- An initial message of the agent is an *initial node* of the automaton.
- Actions of the agent process are *transitions* of the automaton.
- The automaton is Mealy-style [23]; the labels of the transitions in the automaton have the form extracted from actions; an IMDS action $(m,p)\lambda(m',p')$ is converted to a transition from $m$ to $m'$ with a label $p/p'$ ($p$ is an *input symbol* conditioning the transition while $p'$ is an *output*

*symbol* produced on the transition; servers' states are $p=(s,v)$, $p'=(s,v')$).
- For an agent-terminating action $(m,p)\lambda(p')$, a special node *t* in the automaton is added as *target node*, and the transition is of the form $(m,p/p',t)$. For *t* no outgoing transition is defined.
- The system is equipped with a *global input vector* (the vector of global current input symbols), corresponding to a vector of current states of the servers. Firing a transition $(m,p/p',m')$ in the automaton exchanges the symbol *p* with the symbol *p'* in the vector. An *initial global input vector* consists of initial states of all servers.

We denote automaton corresponding to an agent *a* with *ɒ* (reflected *a*). The set of all agent automata we denote as *Ʉ* (reflected *A*, rounded to distinguish it from general quantifier). A *global position* of *Ʉ* is a set of current nodes of agent automata (pending messages of non-terminated agents), and current value (content) of *global input vector*.

The execution of a transition in an agent automaton moves its current node to a target node of the transition (exchanges a message with the output message of the action), and replaces the state of server appointed by the transition to the output state of the action. As in IMDS, it is executed in interleaving manner, and in a case of many transitions possible, the choice is nondeterministic.

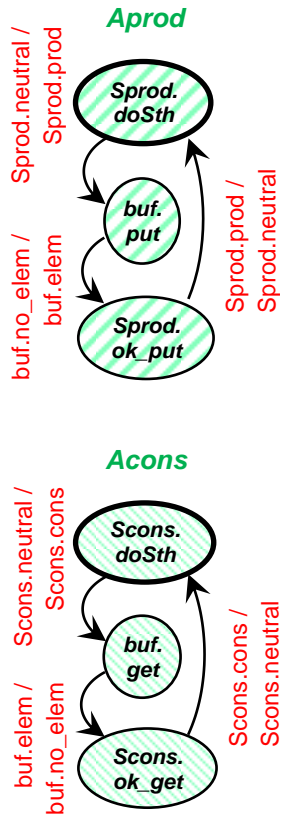

Fig. 5 Agent automata of the *buffer* system

Distributed agent automata for the buffer system are illustrated in Fig. 5. The initial messages of the agents are in bold ovals. Agent identifies are omitted in message labels (nodes of the automata), because they are identical for all messages in a given agent automaton. For completeness, a global input vector of current states of servers should be added.

A global graph of *Ʉ* may be elaborated analogously to the global graph of *Ⱬ*: nodes of the global graph are global positions of *Ʉ*, and edges are transitions in individual agent automata. This graph is analogous to the global graph of $SDA^3$ and to the LTS of IMDS system (global positions contain messages of all non-terminated agents and a vector of states of all servers).

## VI. USING PETRI NET ANALYSIS AND $DA^3$ IN DEDAN PROGRAM

The basic form used in Dedan program is IMDS, because it allows for automatic conversion between the server view and the agent view of a system. Yet, the specification in the form of a relation between pairs $(m,p)\lambda(m',p')$ is exotic for the users. Therefore, an alternative input form of $DA^3$ automata is provided as distributed automata.

A system may be simulated over the global space of configurations (LTS), but it is also possible to simulate it in terms of $SDA^3$. This simulation does not require calculation of a global configuration space of verified system. All of the automata in the system are displayed, with input sets of pending messages under automata identifiers shown. The current states of the automata are distinguished by a separate color.

A user can choose an automaton, and then a list of transitions from the current node of the chosen automaton is displayed (with enabled ones distinguished; it is only one transition in this case, and it is enabled). Next, the user may choose a transition from the enabled ones. If the user clicks an enabled transition, it is "executed" and a destination automaton of the message becomes current.

The internal Dedan verifier, based on CBS evaluation algorithm [24], is limited to deadlock and termination detection formulas. However, the nature of model checking (evaluation of temporal formulas) allows to find only one deadlock in single verification (typically one which generates a shortest counterexample). A user may export a model to the Charlie program which finds many possible deadlocks as elementary siphons. Then, using an external verifier (Uppaal, for example) the reachability of siphon emptying may be examined. The configurations that terminate the counterexamples for every emptied siphon allow to reduce the results, as many siphons may denote the same deadlock [7].

## VII. CONCLUSIONS AND FURTHER WORK

The Dedan program supports an engineer in verification of distributed systems for deadlock freeness, without any knowledge on temporal logics and model checking technique. If a communication deadlock occurs, a sequence diagram of messages is generated, leading from the initial configuration to the deadlock. In a case of resource deadlock, a sequence diagram of changes of servers' states and messages inside agents is generated. If the deadlock is not total, the servers/agents taking part in the deadlock are shown. With Petri net interpretation, some structural properties of a verified system are discovered using Charlie[18]. Distributed automata (in SDA³ or in ADA³ version) allow to design the system in graphical form, and to simulate the components of the system and their cooperation instead of a simulation over the full configuration graph (LTS). Engineers are familiar with the notion of automata (SDA³ are similar to Message Passing Automata [15] and ADA³ are like Timed Automata with global variables of Uppaal [17]) and they may be naturally used in distributed system design. It should be emphasized that simulation over $DA^3$ does not require calculation of global reachability space of a verified system.

TABLE I.
VERIFICATION FACILITIES IN THE THREE EQUIVALENT FORMALISMS

| Formalism: | IMDS | Petri net | DA³ |
|---|---|---|---|
| **Main features** | Specification, model checking, simulation | Structural properties | Graphical input, simulation |
| **Notions** | state | "red" place | • node (S-DA³)<br>• element of global input vector, input/output symbols on transitions (A-DA³) |
|  | message | "green" place | • element of input set (S-DA³), input/output symbols on transitions<br>• node (A-DA³) |
|  | configuration | marking | global position |
|  | action | transition | transition |
|  | initial state | token in red place in initial marking | • initial node (S-DA³)<br>• initial element of global input vector (A-DA³) |
|  | initial message | token in green place in initial marking | • initial element of input set (S-DA³)<br>• initial node (A-DA³) |
|  | initial configuration | initial marking | • initial nodes and initial input sets of all automata (S-DA³)<br>• initial nodes and initial global input vector (A-DA³) |
|  | Labeled Transition System | Marking reachability graph | Global graph:<br>• all states and messages in global positions, input and output symbols (messages) on transitions (S-DA³)<br>• all messages and global input vector in global positions, input and output symbols (states) on transitions (A-DA³) |
| **Features** | • Resource deadlock<br>• Communication deadlock<br>• Partial deadlock<br>• Total deadlock<br>• Partial distributed termination<br>• Total distributed termination<br>• Counterexamples/ witnesses<br>• Configuration space inspection<br>• Simulation over configuration space | Structural properties<br>• Many deadlocks<br>• Existence of siphons<br>• Unreachable actions<br>• Separated components<br>• Invariants | • Graphical definition of a system<br>• Simulation over individual automata<br>• Counterexample projected onto individual automata<br>• Counterexample-guided simulation |

The three specification formalisms: IMDS, Petri nets with restricted structure and distributed automata $DA^3$ (in two forms) are equivalent. The equivalence lays in mapping the basic structures of Petri net and $DA^3$ onto structures of IMDS. All three formalisms generate analogous reachability graphs: LTS for IMDS, reachable markings graph for Petri net and global graph space for distributed automata of both forms. The construction elements and verification features, wider than deadlock and termination identification, are collected for every of the formalisms in the Table I. This variety of specification modes greatly facilitates analysis of the features of distributed systems.

For example, some models of transport cases were modeled. Observation of the server view is equivalent to exchange of messages between road marker controllers that automatically lead the vehicles on the roads. In the agent view, it is the observation of vehicles moving over the road, with interactions to other vehicles occupying some segments of the road. Possible deadlocks in communication may by easily identified, and the verifier shows the behavior of vehicles leading to a deadlock as transitions of $DA^3$ automata. An example may be found in [25]. Server view and server automata allow to observe the system behavior from the point of view of road segment controllers, while agent view and agent automata from the point of view of guided vehicles. Petri net static analysis allows to locate the two possible deadlocks in specification. Such an approach of cooperation of distributed controllers by means of simple negotiation protocols follows the IoT paradigm [26].

The next steps of development of the Dedan environment are:

- *Probabilistic automata* allowing to identify a probability of a deadlock if the alternative actions in system processes are equipped with probabilities.
- *Language-based input* – elaboration of two languages for distributed systems specification: one for the server view (exploiting locality in servers and message passing) and the other one for the agent view (exploiting travelling of agents and resource sharing in distributed environment); a preliminary version of a declarative language-based preprocessor for a server view of verified systems is developed by the students of ICS, WUT under supervision of the author [27].
- *Agent's own actions* – equipping the agents with their own sets of actions, carried in their "backpacks", parametrizing their behavior; this will allow for modelling of mobile agents (agents carrying their own actions model code mobility) and to avoid many server types in specification, differing slightly.

The Dedan environment is successfully used in operating systems laboratory in ICS, WUT. The students verify their solutions of synchronization problems. More than 200 solutions were verified, ranging from several actions to over 5000 actions in a model. Also, Karlsruhe Production Cell benchmark [28] was modeled and successfully verified using the Dedan environment [29]. Several examples of IMDS specification in Dedan input form may found in [30].

In a design process, methods for its automatization are much needed to speed up and make it more dependable. Automatic verification in various forms and automatic code generation [31], which are subject of the research in ICS, WUT, are examples of such trends.